# A phenomenon of AI-conformity: how algorithms change human moral decision-making

Yana Venerina, Dmitry Koch, Nare Meloyan, Gerda Prutko, Valeriia Lelik, Victoria Taova, Andrey Kurpatov

Neuroscience Laboratory, Sberbank, Moscow, Russia

**Abstract**

Social conformity is a well-documented phenomenon in which individuals shift their opinions towards those of a social majority. As artificial intelligence (AI) becomes increasingly integrated into everyday life it may also create a novel source of influence giving rise to algorithmic conformity, mechanisms of which are poorly understood. The present study examined whether AI judgements affect moral decision-making in humans (n=165) adapting the classical Asch paradigm. Participants completed a series of moral dilemmas under three different conditions: in presence of social majority, with an AI model providing brief answers and with an AI model providing both answers and explanations of its choices. In all conditions the presented responses contradicted generally accepted moral norms. The results indicated that an AI model with a reasoning component affected the opinion of participants to a degree comparable to that of a human majority. These findings suggest that even moral judgements, despite their sensitivity and personal significance, may be susceptible to algorithmic conformity. However, the mechanism underlying algorithmic conformity appears to differ from the social one. Overall, the study challenges the assumption that moral decision-making lies in "AI inadmissibility zone" – a sphere that is considered as an area in which only human-made decisions are acceptable and highlights the need for a further investigation of this phenomenon as AI-based recommendations become increasingly embedded into human decision-making.

**Key words:** *algorithmic conformity, AI, moral decision-making*

The authors have no relevant financial or personal conflict of interests.

**Corresponding author e-mail:** yana.venerina@gmail.com

## 1. Introduction

Human judgments rarely occur in isolation from social contexts. Individuals make decisions by balancing their personal convictions with external input from others. Within the group dynamics, people may adjust their decisions towards the group even though this action might contradict their initial decision – a phenomenon known as conformity behaviour (Asch, 1956; Cialdini & Goldstein, 2004).

Social conformity has been systematically investigated for decades, demonstrating consistent effects across various replications, extensions, and meta-analyses (Bond, 2005; Chen et al., 2023; Huang et al., 2023). The core mechanism of this effect stems from two psychological pathways (Deutsch & Gerard, 1955). The first pathway is normative influence, implying that an individual is driven by desires for social acceptance and fear of rejection from a group (Baumeister & Leary, 1995). The second one is information influence based on the assumption that a group's decisions provide more accurate information. Furthermore, these patterns of conformity persist even in online interactions (Wijenayake & Goncalves, 2024). Together, these pathways emphasize that conformity extends beyond behavioural compliance and involves a complex interplay of social-emotional and cognitive factors (Cialdini & Goldstein, 2004).

These findings could alternatively be examined through the view of a system-evolutionary approach (Shvyrkov, 1990; Alexandrov et al., 1997, 2017), that shifts focus from separate psychological processes towards a system framework that considers consciousness and emotions to be different levels of the systemic organization of behaviour. Behaviour is performed to reach an adaptive result – a specific organism-environment relation. Achievement of each result is provided by the activity of various elements of the organism, that we call a functional system (Anokhin, 1974). Systems that were formed in the early stages of the organism development (old systems) are characterized with more emotional patterns, while later stages of development (new systems) are characterized with increase of consciousness components (Alexandrov, Sams, 2005). According to this theory, behaviour is the realization of the history of behavioral development (Alexandrov, Sams, 2005) with different relative prominence of emotional and consciousness components in other words – the activity of old and new systems that provide current behaviour. In that case, we can see conformity as a behavioural strategy that is performed mostly through evolutionary old systems.

In the traditional viewpoint this kind of decision involves not only cognitive evaluation but also empathy and weighing different ethical principles against each other (Greene, 2014). In the frames of system-evolutionary approach we may again say that moral judgment is a complex behaviour that involves old and new systems in the current domain of the individual experience (Alexandrov et al., 2020). In that case, moral judgments under social pressure are supplied by higher contributions of old systems with actualization of a large domain of the individual experience. For instance, previous research has demonstrated that social inference can change moral judgements leading to adjustments of ethical positions under a group's pressure even though these adjustments conflict with their initial moral decision (Kundu & Cummins, 2013). Thus, even fundamental decisions that guide human behaviour can be adjusted.

Over the past five years the rapid development of artificial intelligence (AI) has introduced a novel and powerful source of influence, which renewed interest in the mechanisms of conformity. Since AI systems are capable of inferring patterns and synthesizing new knowledge, they are becoming integrated into the various domains of human life, such as economic decision-making, criminal sentencing, elections, and even rapid decision-making situations such as military operations (Binns et al., 2018; Hackenburg et al., 2025; Lin et al., 2025; Rahwan et al., 2019). Integration of AI into everyday life has led to the emergence of a new research phenomenon – algorithmic conformity – which captures changes in human decision-making under the influence of AI-based systems (Kramer et al., 2018; Liel & Zalmanson, 2025).

Although the observable behavioural outcome of the algorithmic conformity seems to mirror the well-studied effect of social conformity, the underlying mechanisms may differ in several theoretical predispositions. Research on algorithmic conformity suggests that individuals defer to AI recommendations based on beliefs about computational objectivity and superiority (Logg et al., 2019; Dietvorst et al., 2015). From this perspective, AI influence may operate through relatively new systems with higher activation of consciousness components. On the other hand, human advisors are not free from emotional bias, social pressure, or even self-interest. For this reason, AI recommendations are perceived as more persuasive than human ones in contexts where computational analysis is valued (Logg et al., 2019).

Nonetheless, this purely cognitive approach might be incomplete. The Computers Are Social Actors (CASA) paradigm (Nass & Moon, 2000) posits that individuals may apply social rules and heuristics, that originally evolved for interpersonal communications, to AI systems (Xu et al., 2022). If we assume that CASA principles can be extended to modern AI models, then algorithmic recommendations might trigger normative influence similar to human social pressure. Therefore, people may conform not only because they believe that AI is correct but because they experience something analogous to social approval dynamics in human-AI interaction. It is important to note that contemporary AI models differ significantly from the early agents studied under the CASA paradigms. Whereas earlier systems simulated social cues through simple interactions, modern AI models can produce autonomous judgements, engage in forms of moral reasoning, and perform tasks that were traditionally reserved for human deliberation (Rahwan et al., 2019). This significant advancement has prompted a re-evaluation of the CASA framework, conceptualizing AI models as pseudo-social actors (Brandtzaeg et al., 2025; Obrenovic et al., 2025) – an entity capable of eliciting social responses without possessing genuine intentionality or moral agency. Individuals may conform to algorithmic recommendations because they perceive this system as objective and superior compared to themselves (Logg et al., 2019). At the same time, the absence of genuine social presence may exert influence on the emotional and moral significance of conformity and aforementioned effect may elicit unique cognitive and affective reactions distinct from those observed in human groups.

The persuasive power of AI models may be mediated by the provision of explanatory reasoning with recommendations. It has been recently demonstrated that transparency in algorithmic decision-making affects human trust and reliance (Miller, 2019), even though the effects are context-dependent and counterintuitive. Also, explanations can increase trust when they align with user expectations and provide actionable insight (Lai & Tan, 2019). At the same time, they may decrease trust when explanations reveal algorithmic limitations or reasoning that conflicts with human values (Dodge et al., 2019). In the context of moral decisions, AI explanations may cause the following consequences: (i) they provide information that can be evaluated against one's own moral reasoning; (ii) they show transparency; (iii) they may anthropomorphize the AI model by demonstrating a capacity for reasoning that is usually associated with humans. Therefore, explanation may enhance perception of AI as a legitimate reasoner, leading to a significant increase in conformity beyond what simple recommendations achieve. Explanations might emphasize the algorithmic nature of AI judgements and reduce their persuasiveness where empathy and contextual sensitivity are deemed essential.

The influence of AI recommendations on individuals' judgements has shown controversial results. For example, Logg and colleagues (2019) presented an effect of algorithm appreciation that implies a tendency to forego human predictions for algorithmic ones in objective tasks. However, Dietvorst et al. (2015) demonstrated an effect of algorithm aversion that is an opposite pattern to the previous one. In this case, participants avoided algorithmic advice even though it outperformed humans' predictions on average. This aversion effect is more pronounced in tasks that require subjective judgement or moral evaluation (Castello et al., 2019). Recent studies have shown similar controversial results. While Kramer et al. (2018) found that social robots could induce conformity in perceptual tasks at rates almost reaching human confederates, Wijenayeke and Goncalves (2024) presented differences between online human influence and algorithmic recommendations. Yet, their studies examined either factual or perceptual judgements rather than moral decisions. Therefore, the question of whether AI can exert comparable influence in moral decisions is still under consideration. Furthermore, previous studies have not varied the explanatory transparency of AI recommendations to determine whether provided reasoning may alter their persuasive effect relative to human consensus.

The current study aimed to address these gaps through an experimental manipulation that allowed us to compare AI (algorithmic) conformity vs. human conformity in moral decision-making and to assess the impact of decision source type on conformity level. We selected moral dilemmas based on three theoretical considerations. First, moral decisions engage both cognitive and emotional processing, allowing us to investigate whether AI influence extends beyond computational tasks. Second, moral contexts are more ambiguous than cognitive or perceptual tasks that might lead to a different outcome regardless of AI's objectivity. Third, the increasing integration of AI makes understanding AI influence on moral judgements practically important (Ram, 2025). The study provided an important opportunity to advance the understanding of algorithmic conformity and its application in moral judgements.

## 2. Methods

### 2.1 *Participants*

The experiment involved 165 participants (50% women, age M = 28.4 , SD = 8.04). All participants had normal or corrected-to-normal vision and had no history of neurological or psychiatric disorders. We used a between-subject design in which participants were randomly assigned into the control and the three experimental groups: (i) the AI group, (ii) the AI reasoning group, and (iii) the Human group. Individuals were recruited through a professional research agency which was responsible for their screening, scheduling, and compensation. Each individual received a monetary reward compensating for their time from the agency upon completion of the experimental sessions. The study protocol (№ 4, 24.11.2025) was reviewed and approved by the Independent Ethics Committee of the Artificial Intelligence Research Institute (AIRI), and all participants signed the informed consent form prior to the experiment. None of the participants reported prior experience in conformity or deception-based experiments.

### 2.2 *Experimental Design*

All participants passed through the Russian adaptation of Analysis-Holism Scale questionnaire (Choi et al., 2007; Apanovich et al., 2017) to take to account culture specific aspects of moral decision making, and to create answers, that would be mostly opposite to the moral percepts of our respondents. Moreover, the control group was recruited to check typical answers to the suggested dilemmas for the same reason.

The experimental group was divided further into three groups: the Human group, the AI group, and the AI reasoning group. The experimental task for each participant was to respond regarding eighteen presented moral dilemmas on the broadcast screen. Although the experimental task was equal between conditions, the source of influence varied between groups. In the control condition, participants responded regarding moral dilemmas without any social pressure or influence of the AI model. The results of this condition were further used as a reference point for experimental conditions.

### 2.3 *Materials and Procedure*

The experimental procedure began with social questionnaires. After filling in all necessary information, a participant had to respond to the PANAS questionnaire. Further, participants were presented with moral dilemmas adapted from the "trolley problem" family. Each scenario required a binary decision (e.g., whether to sacrifice one life to save several others) (see Appendix 2). Upon completion of the experimental task, participants completed the PANAS and NASA-TLX questionnaires. We hypothesized that internal conflict arises when individuals' own choices contradict external recommendations from either a group or an AI model. This contradiction may induce internal tension, manifested through changes in affect and subjective perception of effort required for decision-making. The scenario of answers contained deontological responses that contradicts less utilitarian culture-specific patterns of the Russian population (Arutyunova , et al., 2016) and opposes the answers of the control group. Further, we provide a thorough description of experimental manipulation.

#### *2.3.1 Human Majority Procedure*

In the Human condition, conformity pressure was created through a simulated group discussion conducted via an online video call. Each experimental session included one genuine participant and six confederates who appeared as other participants. All individuals' cameras and microphones were active, and everyone responded verbally to each task in a fixed order. The confederates were instructed to provide identical predetermined answers on each trial, regardless of correctness. The real participant was always invited to answer third from last to ensure consistent exposure to the unanimous opinions of the group before giving their own response. This setup reproduced the key structure of Asch's conformity paradigm in a naturalistic digital environment, creating authentic social pressure while maintaining full experimental control. In each trial participant had to respond verbally and then to put his answer into an individual form. The experimenter moderated the session, ensuring smooth turn-taking and a neutral communication tone. After completing all tasks, the naïve participant was fully debriefed and informed that the other participants had been confederates instructed to follow a scripted response pattern.

#### *2.3.2 Wizard-of-Oz Procedure*

To simulate interaction with an AI system, we employed a Wizard-of-Oz setup involving a confederate who operated the experimental interface without participants knowledge. Participants were informed that an autonomous AI system was taking part in the same task, independently analyzing the stimuli and providing its own responses. In reality, the "AI" output was generated through prerecorded audio messages voiced in a synthesized, robotic tone to strengthen the perception of machine agency. The confederate triggered specific audio responses corresponding to each experimental condition ("AI" or "AI with Reasoning") according to a predetermined sequence. In the AI with Reasoning condition, the pre-recorded audio messages not only presented a decision but also provided a brief philosophical justification of that choice. Each message contained one or two sentences referencing ethical principles or normative frameworks – for example, "From the perspective of harm minimization, option B preserves more lives," or "According to the principle of double effect, the moral permissibility of this action depends on intent rather than outcome." This procedure allowed precise control over the timing and content of the AI's outputs while preserving participants' belief that they were observing the behaviour of an autonomous analytical system. In each trial participant had to respond verbally after the AI system and then to put his answer into an individual form.

Participants were fully debriefed after the experiment and informed that the AI interaction had been simulated.

### *2.4 Psychological Measures*

A combination of quantitative, qualitative, and psycholinguistic measures was employed to capture behavioural, cognitive, and emotional aspects of conformity across experimental conditions.

Based on the diagnostics of analytic-holistic cognitive style (Apanovich et al., 2017), the majority of respondents (N = 124) showed a pronounced shift toward the holistic processing strategy. The analytic style was identified in 8 respondents (6.5%), whereas the holistic style was characteristic of 116 respondents (93.5%), allowing us to proceed our analysis further.

The primary behavioural indicator was the conformity rate, calculated as the proportion of conformist responses across all critical trials. The experimental task consisted of 18 items in total, of which 6 were control and 12 were critical. The order of trials was randomized at the design stage, but then kept fixed across participants within each experimental scenario. On control trials, the majority (or AI) provided socially desirable or normatively correct answers, which served as a baseline for participants' spontaneous judgements. On critical trials, however, the majority (or AI) intentionally selected socially undesirable or counter-normative options. A participant's agreement with these counter-normative responses was coded as an act of conformity. The conformity rate thus represented the proportion of socially undesirable answers that coincided with the majority or AI choice, providing a direct quantitative estimate of susceptibility to external influence under each experimental condition.

To assess participants' cognitive load, the NASA Task Load Index (NASA–TLX, adapted by Rybina et al., 2023) was administered following the task series. Participants were asked to evaluate how much mental effort they felt was required to complete the task using a continuous 0–100 scale, where 0 indicated "Low load" and 100 indicated "High load." The instruction presented on screen stated, "Please rate how much mental effort you felt was required to complete the task. Provide a score from 0 to 100, where 0 indicates low and 100 indicates high workload." This measure yielded subscale scores for mental demand, effort, and frustration, as well as a composite workload index, enabling comparison across task types and conditions.

Participants' affective states were measured with the Positive and Negative Affect Schedule (PANAS, adapted by Osin, 2012) twice – before the experimental session and immediately after the experimental block. The PANAS consists of a list of adjectives describing emotional states, and participants were instructed to indicate to what extent they felt each emotion "at the present moment" on a 5-point Likert scale ranging from 1 ("Not at all") to 5 ("Very much"). Separate mean scores were computed for positive and negative affect, allowing us to evaluate how different sources of pressure – social or algorithmic – shaped the participants' immediate emotional responses.

### *2.5. Post-assessment interview*

Beyond these quantitative measures, we collected qualitative data through semi-structured post-experimental interviews. After completing both task series, 60 randomly chosen participants from each experimental group (n = 20, 50% of women) were invited to reflect on their reasoning processes, emotional reactions, and perceptions of the source of influence. The interviews followed a flexible guide that probed how participants explained their decisions, whether they trusted or resisted the AI, and what emotions they experienced when disagreeing with it. All interviews were recorded and transcribed. After the procedure 3 interviews (one from each group) were excluded from analysis because of poor quality of audio files, so the balance of the groups wasn't disturbed. Interviews are included in the section Results as illustrative material to enrich the interpretation of the behavioural findings, rather than self-standing outcomes of systematic analysis.

### *2.6 Statistical analysis*

To assess the degree of both social and algorithmic conformities among participants when resolving ethical dilemmas, we calculated an index of mean absolute deviations (delta) from the “predicted” responses. These baseline (“predicted”) responses were derived from the control group and served as a reference point for estimating conformity in the experimental conditions. Lower values of this index indicate greater agreement with the “predicted” solution, reflecting higher levels of conformity. On the other hand, the higher values indicate lower agreement and conformity level. We run a one-way ANOVA with Welch’s correction to test the difference between experimental conditions. Post-hoc comparisons for behavioural data were conducted by the Games-Howell correction.

## 3. Results

### *3.1. Behavioural Results*

To examine the effect of decision source type on conformity level a one-way ANOVA with Welch’s correction was conducted. The analysis revealed a statistically significant main effect of decision source type on conformity ($F(2, 77.6) = 9.56$, $p < .001$, $\eta^2_p = 0.13$).

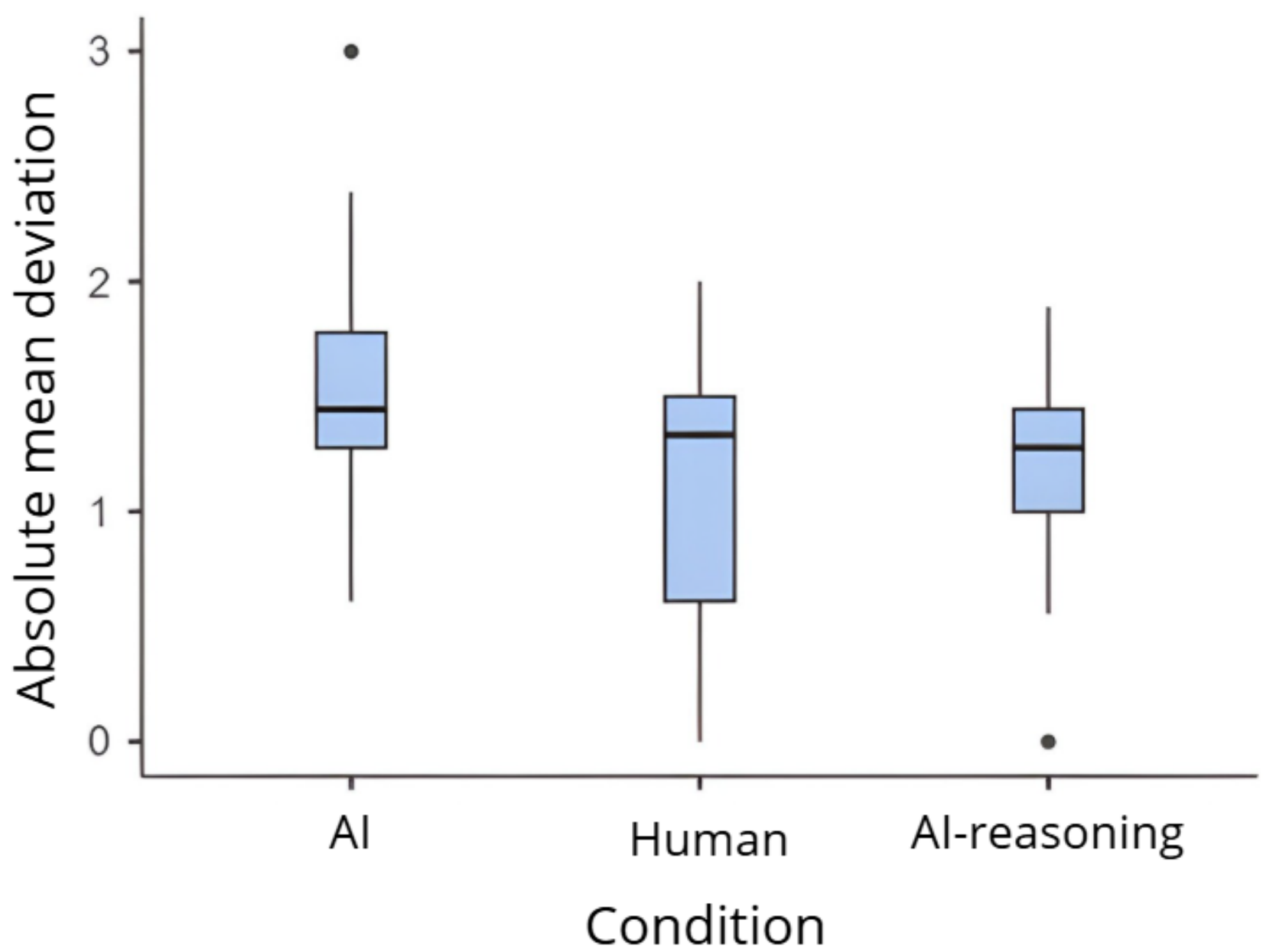

**Figure 1.** The results of ANOVAs between experimental conditions.

Individuals in the AI-reasoning group (Mean = 1.23, SD = 0.37) demonstrated higher conformity compared to the AI group (Mean = 1.56, SD = 0.37, $t = 3.68$ $p = .001$, Cohen’s $d = .80$), whereas it was lower compared to the human group (Mean = 1.12, SD = 0.598, $t = 3.75$, $p < .001$, Cohen’s $d = .83$). Notably, conformity levels in the AI-reasoning group did not differ from the human group (M = 1.12, SD = 0.59, $t = 0.96$ $p = .60$, Cohen’s $d = .22$), suggesting functional equivalence between explained AI recommendations and human influence. Results of post-hoc comparisons using the Games-Howell test are presented in Table 1 in Supp and Figure 1).

Important to note a striking behavioural dissociation was observed in the Human group 56.15% of verbal answers of the Human group participants dramatically differed from the answers that were put into their individual forms that they fulfilled in parallel with the experimental procedure, comparing to 2.6% in the AI Reasoning and 0.8% in the AI groups. This effect gives us an opportunity to conclude that the answers we provide as reference were really contradictory to the personal opinion of most of the participants in the Human group and caused internal controversy.

### *3.2. Results of Questionnaires*

To ensure that exposure to external recommendations creates internal conflict over individuals during decision-making, we analyzed indicators of subjective cognitive load (NASA-TLX) and emotional state (PANAS) before and after the experiment. We hypothesized that external recommendation might induce internal tension, manifested through changes in affective and subjective perception of decision-making effort. Furthermore, we assume this effect would be more pronounced for the AI reasoning group.

*Emotional responses (PANAS results)*

A mixed-design ANOVA was conducted with Time (pre/post) as the within-subjects factor and Condition (AI group, AI reasoning group, Human group, control) as the between-subjects factor. Post-hoc comparisons were conducted by the Games-Howell test.

*Positive Affect*

The analysis revealed a significant main effect of Time on positive affect, $F(1, 161) = 12.61$, $p < .001$, $\eta^2_p = .073$, indicating that positive affect significantly decreased following the moral dilemma task (MD = -1.63, SE = 0.46, $t = 3.55$, $p < .001$, Cohen's $d = 0.28$). However, neither the main effect of Condition, $F(3, 161) = 0.29$, $p = .831$, $\eta^2_p = .005$, nor the Time × Condition interaction, $F(3, 161) = 0.77$, $p = .510$, $\eta^2_p = .014$, reached statistical significance. Post-hoc comparisons revealed no significant pairwise differences between conditions ($p$s > .78), indicating that the decline in positive effect was uniform across all experimental groups regardless of decision source type.

*Negative Affect*

Similarly, the analysis of negative affect showed a significant main effect of Time, $F(1, 161) = 10.74$, $p = .001$, $\eta^2_p = .063$, with participants reporting increased negative emotions following the task (MD = 2.17, SE = 0.66, $t = 3.28$, $p = .001$, Cohen's $d = 0.26$). The main effect of Condition, $F(3, 161) = 0.94$, $p = .425$, $\eta^2_p = .017$, and the Time × Condition interaction, $F(3, 161) = 0.93$, $p = .427$, $\eta^2_p = .017$, were non-significant. Post-hoc comparisons confirmed no significant differences across condition pairs ($p$s > .43), demonstrating that the increase in negative affect was consistent across all groups.

*Cognitive Load*

To examine differences in subjective cognitive load across decision source conditions a one-way ANOVA was conducted. The analysis revealed no statistically significant differences in perceived cognitive load across conditions, $F(2, 80) = 1.60$, $p = .303$, $\eta^2_p = .029$. Post-hoc comparisons using the Games-Howell test confirmed no significant pairwise differences across conditions ($p$s > .31; see Table 2). These findings indicate that subjective cognitive load remained stable across all experimental groups, suggesting that the type of decision source (Human, AI, AI reasoning) did not influence participants' perception of task difficulty or required effort.

### *3.3. Post-assessment Interview Analysis*

Following the task completion, participants were then asked to discuss the task and their own perception of moral decision-making (see Supplementary for original sentences).

Q1: How did you perceive the group/AI model throughout the course of the experiment? To what extent did it seem confident or exhibit expertise in its answers?

Most of the participants in the Human group tend to appeal to the individual experience of the social group members, pointing out that morality is a product of social nurture that everybody receives. On the other hand, some of them expressed doubts regarding the validity of the moral stances, stemming from a lack of relevant life experience.

*'I think we are all equally informed about morality – we are all people who were raised [...] and lived in society' - participant 133, the Human group*

*'Well, at first glance it is hard to tell how informed a person is on matters of morality [...] if there were, I do not know, a priest in a robe, then I would certainly think about it [...] You have to go by a person's upbringing [...] Most people have never actually thought over such questions. [...] Those who have, based it on some books and films with happy endings [...] (they see the world) through rose-colored glasses [...] That is not usually the case.' - participant 32, the Human group.*

Respondents pointed to a perceived logical superiority and a higher degree of ethical literacy within the AI reasoning model. While its rationality was regarded as a benefit in situations necessitating swift and maximally effective resolutions, participants simultaneously emphasised the model's absence of empathy and self-sacrificial capacity needed for moral judgements.

*'He put forward arguments about moral and ethical norms, I agree. [...] He probably has all these moral norms uploaded into him, and he is unable to deviate (the answer) so much as to step to the left or right [...] He is able to assess a situation faster, I would say'. - participant 73, the AI reasoning group.*

*'The fundamental postulates are built into him [...] He draws conclusions based on what has been put into him. [...] May you consider him more knowledgeable about morality (moral questions) than yourself? – Yes.' - participant 75, AI reasoning group.*

At the same AI group of respondents highlighted the tyrannical and ruthless nature of the AI's decisions, while simultaneously speculating that the model may have been under-trained.

*"It seems to me that it is simply a soulless machine that mathematically calculates how many people it is more advantageous to save, for example. Disregarding any thought to the fact that one person was innocent and had the right to live. Well, it is just a machine – it reasons either rationally or from a legal standpoint. And maybe there are instructions for such unforeseen situations, and it simply follows those instructions. Or perhaps there is the instruction where it has to save as many people as possible, and that is what it has done.' - participant 54, the AI group.*

*"[My opinion] diverged, I think, but it (the AI model) has not been fully trained yet. That was the only idea I had – that there is room for improvement.' - participant 55, the AI group.*

*"As I understand it, AI has not yet been developed to a sufficient degree to be trusted on questions like these" - participant 60, the AI group.*

Q2: Recall the moment you first became aware that your initial judgement regarding a moral decision in the task diverged from that of the group/AI model. Please describe one such instance that stands out in your memory

Respondents of the Human group aligned their decisions with the constructs of humanness and empathy. Being set in opposition to the group likely appealed to low-differentiated systems and prompted reflection on one's alignment with this group based on the criteria of humanity.

*"At that moment I was wondering whether my position was rational and not cruel' - participant 164, the Human group.*

Interaction with the reasoning AI model led to an attempt to adopt a more rational stance in moral decision-making.

*'(AI) it led me to think as if it were murder." - participant 73, the AI reasoning group.*

*"[...] I considered it categorically wrong, but the AI added that the situation was ambiguous because all people had been in the same position, and the right position could be to save five people". - participant 76, the AI reasoning group*

The AI model was not regarded by respondents of the AI group as a definitive moral or rational reference point. In cases of divergence, they tended to believe the model was in error due to being under-trained.

*"Listening to a completely opposite answer from the AI, I wonder ... perhaps, I had not understood the wording of the questions correctly. But then, after re-reading the question, I realized that my answer was, after all, the more moral and appropriate one. [This is about] the scenario with the doctor" - participant 58, the AI group.*

Q3: Did you feel that the final response in these disputed instances was entirely your own decision? Alternatively, did it feel as though the answer had been 'prompted' by the group/AI model? In what way did this sense of authorship — or the absence of it — manifest itself?

Respondents of the Human group described their experience through their emotional states, which may reflect the actualisation of low-differentiated experience. Their accounts included negative impressions associated with the perceived compulsion to conform to the majority opinion or to resist it.

*"Well, drawing on my own experience and life situations, I am going to stay by my choice. [...] Perhaps this feeling of frustration led me to the agreement with the majority' - participant 129, the Human group.*

*"I felt this pressure of a divergent public opinion [...] anxiety is getting higher, fear, a kind of disappointment - perhaps in myself [...] because I could not fit into the world" participant 30, the Human group.*

*"Perhaps fear of judgement. [...] That there would be some kind of social punishment. [...] There was a sense of pressure, of tension [...]" - participant 32, the Human group.*

At the same time some marked a pleasant sense of unity with the group, which notably was not accompanied by a loss of the sense of agency:

" *[...] the fact that I was following the group felt pleasant as if we aligned in the same direction as society. Second, it felt great that this was still my own decision, one I had thought through due to the group's opinion. but without blindly following the group". participants 32, the Human group.*

Respondents of the AI reasoning group viewed the AI model as a guide for decision-making, perceiving its responses as a steer that enabled them to provide the "correct" answer, yet they did not feel they had relinquished their sense of agency upon following its suggestions.

*"The AI in this case served as a confirmation of whether I was following the right direction. I think it would have been more difficult for me to choose without it." participant 71, the AI reasoning group.*

*"Ultimately, this was my decision because one way or another I believed that position was correct. [...] the AI elucidated it, made it clearer [...] Yes, there was something of a hint. [...] I felt that the authorship was mine - that there was no one beside me and I was making this decision entirely on my own." participant 75, the AI reasoning group.*

Respondents of the AI group have just pointed out the low authority of the AI-model.

*"In my view, AI differs in such a way that it does not understand anything at all. [...] It has no real opinion, no personality, no genuine competence. [...] Besides, what is AI going to do to me? A crowd might judge me. There is a moral dimension of this question. There could be consequences." - participant 50, the AI group*

*"I am superior, my decision is more correct regardless [of AI decisions]. I am unquestionably superior to AI. [...] There is no single right answer, and my decision would be like this even if it is different from an advanced AI." - participant 56, the AI group.*

To summarize, the qualitative data reveal a fundamental divergence in how participants perceive moral authority depending on whether their opponent was human beings or AI. Reasoning of AI played a crucial role in decision-making, shifting the respondents from emotional perception toward more analytic and rational reflection.

## 4. Discussion

As AI systems become increasingly integrated into everyday life, they are emerging as important sources of information and guidance. This study investigated whether AI recommendations influence conformity in moral-decision making and whether explanatory reasoning moderates this effect. The results indicated that AI recommendations accompanied by justifications produced conformity rates similar to human majority influence. Strikingly, exposure to the AI reasoning demonstrated a significantly higher conformity rate in the AI group, achieving levels that are equivalent to those observed under human social pressure.

Importantly, the Human condition showed a significant dissociation between public and private responses. Participants often agreed with the majority verbally but later gave different answers in individual forms. This pattern is consistent with normative social influence rather than genuine internal acceptance, contrasting to the AI and AI reasoning group. Participants of the AI group have just followed their own opinion, and the influence of AI answers wasn't so significant. Participants of the AI reasoning group have genuinely updated their moral judgements rather than responding out of concern for social consequences. This evidence has a strong practical implication: while human social influence produces situation-dependent conformity, AI influence via reasoning may lead to deeper attitudinal change that persists beyond the immediate interaction.

Moreover, our findings reveal differences in how participants perceive their "moral opponents" depending on their nature. While the Human group considered choices of other participants contradicting their opinion based on their social and individual experience, the AI group viewed the same answers of the "AI model" as erroneous and given by a "soulless machine" that was under-trained, at the same time, the AI reasoning group qualified the model as unbiased and rational agent with high degree of "ethical literacy".

Research performed by Krügel et al. (2023) on moral decision-making with ChatGPT advice suggests that people tend to follow the model's recommendations, but what is more concerning, they underestimate ChatGPT's influence on their decisions, believing that they were fully unbiased and autonomous. Our findings extend this line of evidence suggesting that AI recommendations accompanied by reasoning may be perceived as more valid because it is not associated with emotional subjectivity of human judgement. Participants in the AI reasoning group reported higher levels of subjective satisfaction with their answers than those in other conditions, which may indicate that AI-generated reasoning made participants feel that their answers were more logical and justified. This may reflect that the reasoning of AI doesn't arise a controversy, but instead produces an illusion of understanding of the explanatory logic, which may mislead participants away from the correct answer.

We have hypothesised that cognitive load would be higher in the Human group because of additional resources needed to spend to perform social interaction in this condition. Interesting, but contrary to our hypothesis, subjective cognitive load did not differ across conditions. One possible explanation is that decision scenarios were experienced as emotionally intense or affectively charged across all groups, thus decreasing sensitivity of workload ratings. At the same time, the absence of significant differences should be interpreted cautiously, as subjective workload measures may not capture more subtle variations in cognitive processing that do not reach conscious awareness.

According to the system-evolutionary approach, behaviour can be understood as the joint realization of less differentiated emotional systems and more differentiated conscious systems (Alexandrov, Sams, 2005). Within this framework, making decisions together with AI have a limited representation in the structure of subjective experience. From the ontogenetic viewpoint, AI appears in rather late stages of our individual development. Here we speculate that the integration of AI into the moral decision-making process is perceived by an individual as a way to make more balanced and rational decisions as a result of the increased contribution of highly differentiated systems to his behaviour. This creates the basis for making erroneous judgments, since, apparently, it decreases the relative impact of low-differentiated systems of the entire domain of relevant experience leading a human to more analytic, utilitarian decisions. As a result, a person can make unjustified decisions based on AI reasoning. According to the idea of culture-specific cognitive styles that determine moral decision-making as well (Arutyunova et al., 2016), Russian culture is considered as holistic one, with prevalence of deontologic approach in solving dilemmas (Arutyunova et al., 2016). These findings are particularly noteworthy as AI reasoning appears to shift decisions away from this culturally established pattern and toward the opposite tendency.

There is a growing body of studies that highlights the importance of perceived computational objectivity and superior processing capacity in human deference to algorithmic recommendations (Logg et al., 2019; Hou et al., 2021; Jussupow et al., 2024). Research demonstrates that people often utilized AI recommendations as a cognitive shortcut leading them to perceive algorithms as more objective, accurate and free from biases compared to human advisors. This perception drives algorithm appreciation particularly for analytical tasks rather than subjective judgement. However, the results suggest that these factors alone are not sufficient when decisions involve moral decisions. Our findings support previous studies showing that algorithm aversion increases in high-stakes scenarios involving moral considerations, and that people prefer human decision-makers for tasks involving ethical dilemmas or emotional interactions.

Moreover, they may serve as a decision-making boundary: while unexplained AI recommendations were not as effective as other conditions to influence moral judgements, AI recommendations with reasoning achieved influence similar to human social pressure. This suggests that participants integrate argumentation of AI into their own reasoning. This finding is also consistent with the CASA framework, which suggests that AI systems are perceived as objective and superior, contributing to algorithmic conformity (Logg et al., 2019). This interpretation is further supported by participants' responses in the AI reasoning group interviews. Many of them mentioned intellectual superiority of the AI model, pointing to its access to ethical norms of humanity and absence of emotional biases.

We assume that these results might be explained by several complementary mechanisms that together transform AI from a relatively weak into a compelling influence source. First, philosophical justifications provide valuable information that participants can evaluate against their own moral intuitions, functioning as a form of informational influence (Gerard & Deutsch, 1955). When the AI system articulated moral principles, participants may have perceived this information as relevant input into their own reasoning. Recent studies support this interpretation, demonstrating that reasoning increases trust particularly when they provide feature-relevant information (de Brito Duarte et al., 2023). Second, we hypothesise that reasoning may serve as an anthropomorphizing function by demonstrating abilities that are associated with human cognition. This aligns with recent findings that conformity to AI is partially mediated by normative pressure, including the discomfort experienced when disagreeing with AI systems that demonstrate apparent reasoning capacity (Liel & Zalmanson, 2025b). It is possible that this demonstration of reasoning capacity activates social heuristics described in the Computers Are Social Actors paradigm (Nass & Moon, 2000).

It is important to acknowledge several limitations of our study. First, the relatively modest sample size may limit the generalizability of the findings. Replication with larger and more diverse samples would strengthen the current findings. Second, the "Wizard of Oz" paradigm may have introduced challenges in maintaining full standardization of the procedure. The laboratory context may impose constraint due to the observation process itself – participants may have approached the task with an awareness of being observed by researchers. Furthermore, we utilized only 18 moral dilemmas without varying the approach of solving it which can be expanded in further studies.

## Summary

Our findings suggest that reasoned justifications provided by AI models can shape people's moral judgements by promoting the perception of reasoning AI as a purely cognitive, unbiased, and "ethically literate" agent. It carries several significant implications such as inadvertently bypassing individuals' own emotional, cultural and experiential moral systems leading to a deceptive sense of objectivity that masks algorithmic influence.

## Acknowledgements

We would like to thank Polina Chudina for her help in verification of the post-assessment interview analysis, and Dina Popova for productive discussions and critical assessment of the text.

## Supplementary

Table 1. Post hoc comparisons for Behavioural results of conformity between experimental groups.

| **Comparisons** | *ΔM* | *SE* | *p* | *Cohen's d* |
|---|---|---|---|---|
| AI – Human | 0.420 | 0.120 | .001** | 0.83 |
| AI – AI reasoning | 0.314 | 0.087 | .001** | 0.80 |
| AI reasoning – Human | 0.106 | 0.150 | .604 | 0.22 |

*Note.* ΔM — Mean difference; SE — standard error; p-value was corrected by Games–Howell method.

Table 2. Post hoc comparisons for NASA Task Load Index Summary Scores by Condition

| | **Descriptive statistics** | | | | **Games-Howell post hoc comparisons** | | | | |
|---|---|---|---|---|---|---|---|---|---|
| **Condition** | *N* | *M* | *SD* | *SE* | **Comparison** | *MD* | *t* | *df* | *p* |
| AI only | 41 | 247 | 104.0 | 16.2 | AI only vs. Human only | −29.0 | −1.23 | 79.8 | .440 |

| Human only | 41 | 276 | 109.9 | 17.2 | AI only vs. AI reasoning | −31.6 | −1.46 | 79.3 | .314 |
|---|---|---|---|---|---|---|---|---|---|
| AI reasoning | 42 | 279 | 91.9 | 14.2 | Human only vs. AI reasoning | −2.54 | −0.11 | 77.9 | .993 |

**Note.** MD = mean difference.

Appendix 1

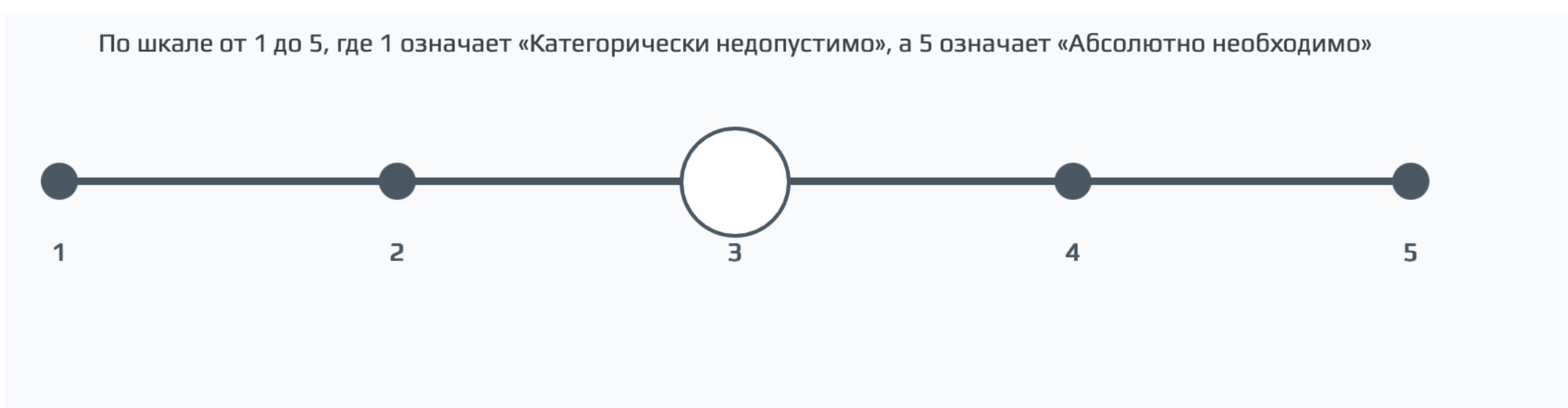

**Please rate on a scale from 1 to 5, where 1 indicates “Completely unacceptable” and 5 indicates “Absolutely necessary.”**

1. Управляя моторной лодкой, Андрей замечает пятерых пловцов, которые тонут вдалеке. Если Андрей не поплывет к ним с максимальной скоростью, он не успеет их спасти, и все пятеро погибнут. Чтобы плыть с максимальной скоростью, Андрей должен резко прибавить газ.
   Если он резко ускорится, то спасет пловцов. Если он не сделает этого, они все утонут. Андрей решает резко прибавить газ.
   Резко прибавить газ – это:

   *Andrei is driving a motorboat when he notices five swimmers drowning in the distance. If Andrei does not head toward them at full speed, he will not reach them in time and all five will die. To go at full speed he must suddenly accelerate. If he accelerates sharply, he will save the swimmers. If he does not, they will all drown. Andrei decides to accelerate sharply. Accelerating sharply is:*

2. Стоя на пешеходном мосту над рельсами, Федя увидел, что пустой неуправляемый вагон вот-вот собьет пять человек. Нога Феди застряла в перилах, но рядом с Федей стоит еще один человек, которого можно толкнуть, чтобы он упал с моста на рельсы и попал под вагон. Из-за этого человека скорость вагона снизится, и он не собьет пять человек. Если Федя столкнет одного человека, тот попадет под вагон, который замедлит ход и пять человек выживут. Если Федя не столкнет этого человека, вагон продолжит движение и собьет пятерых, а один человек останется в живых, стоя на мосту над рельсами. Федя решает столкнуть человека.
   Столкнуть человека – это:

   *Standing on a pedestrian bridge above the tracks, Fedya sees an unpowered, runaway railcar that is about to strike five people. Fedya’s foot is stuck in the railing, but another person is standing next to him; that person could be pushed so that he falls from the bridge onto the tracks and into the path of the car. Because of this person the car’s speed would slow and it would not hit the five people. If Fedya pushes that person, the person will be hit by the car and the five people will survive. If Fedya does not push,*

*the car will continue and hit the five people, while one person on the bridge will remain alive. Fedya decides to push the person.*
*Pushing the person is:*

3. На новой выставке аквариумов Коля видит, как посетитель поскользнулся на мокром полу, упал и сломал себе шею. Он еще жив, и, если останется неподвижным, то его могут безопасно эвакуировать медики. Однако он упал на шланг, который подает кислород пяти другим посетителям, находящимся на подводной экскурсии. Без кислорода эти пять посетителей скоро умрут. Если Коля ничего не сделает, то первого посетителя безопасно эвакуируют, но пять человек, находящихся в батискафе, умрут. Если Коля сдвинет упавшего посетителя со шланга, то он умрет, но пять посетителей в батискафе получат кислород и выживут. Коля решает сдвинуть первого посетителя.
   Сдвинуть посетителя – это:

   *At a new aquarium exhibition, Kolya sees a visitor slip on a wet floor, fall, and break his neck. The visitor is still alive, and if he remains still, medics can safely evacuate him. However, he has fallen on a hose that supplies oxygen to five other visitors on an underwater tour. Without oxygen, those five visitors will soon die. If Kolya does nothing, the first visitor will be safely evacuated but the five in the submersible will die. If Kolya moves the fallen visitor off the hose, that visitor will die, but the five in the submersible will receive oxygen and survive. Kolya decides to move the first visitor.*
   *Moving the visitor off the hose is:*

4. Стоя возле железной дороги, Денис увидел, что пустой неуправляемый вагон вот-вот собьет пять человек. Рядом с Денисом находится стрелочный рычаг, которым можно направить вагон на запасной путь и в сторону от пятерых людей. Но перевод стрелки также приведет к опусканию перил пешеходного моста, находящегося над запасным путем. В результате этого один человек упадет с моста на рельсы запасного пути и попадет под вагон.
   Если Денис потянет за рычаг, вагон пойдет на запасной путь и не собьет пять человек, а один человек упадет на рельсы и попадет под вагон. Если Денис не потянет за рычаг, то вагон продолжит путь, собьет пять человек, а один человек на мосту запасного пути останется жив. Денис решает потянуть за рычаг.
   Потянуть за рычаг – это:

   *Standing by the railway, Denis sees an empty runaway wagon that is about to hit five people. Near Denis is a switch lever that can divert the wagon onto a siding and away from the five people. However, switching the track will also release the railing of a pedestrian bridge over the siding, causing one person to fall from the bridge onto the siding and be hit by the wagon. If Denis pulls the lever, the wagon will go onto the siding and not hit the five people, but one person will fall onto the tracks and be struck by the wagon. If Denis does not pull the lever, the wagon will continue and hit the five*

*people, and the one person on the bridge near the siding will remain alive. Denis decides to pull the lever.*
*Pulling the lever is:*

5. Дима едет по заливу на моторной лодке, как вдруг замечает пловцов, которым грозит опасность. Пятеро пловцов тонут в конце узкого канала прямо по курсу Димы. Сбоку от канала тонет еще один пловец. Если Дима остановится и спасет этого одного пловца сбоку от канала, он не успеет спасти пятерых пловцов. Если Дима прибавит скорость и поплывет к пятерым пловцам мимо одного пловца, последний утонет, но пятерых пловцов удастся спасти. Дима решает прибавить скорость и плыть в сторону пятерых пловцов.
   Прибавить скорость к пятерым пловцам – это:

   *Dima is driving a motorboat through a bay when he notices swimmers in danger. Five swimmers are drowning at the end of a narrow channel directly in Dima's path. To the side of the channel there is one more swimmer who is also drowning. If Dima stops to rescue that single swimmer, he will not be able to save the five. If Dima increases speed and goes past the lone swimmer toward the five, that lone swimmer will drown but the five will be saved. Dima decides to speed up toward the five swimmers.*
   *Speeding toward the five swimmers is:*
6. Рома – пожарник, он пытается спасти пятерых детей из горящего дома. В доме только одно окно, откуда можно безопасно эвакуировать детей, и его заклинило. Рома должен немедленно выбить это большое тяжелое окно, иначе все пять детей погибнут. За окном на подоконнике находится человек, в безопасности ожидающий эвакуации, которого Рома неизбежно столкнет с подоконника, если выбьет стекло. Падение с подоконника несомненно окажется для него смертельным. Если Рома выломает окно, он собьет человека, и тот погибнет, упав с подоконника, но пятерых детей удастся эвакуировать. Если Рома не выломает окно, то человека спасут, но погибнут дети. Рома решает выломать окно.
   Выломать окно – это:

   *Roma is a firefighter trying to save five children from a burning house. There is only one window through which the children can be safely evacuated, and it is jammed. Roma must immediately break out this large, heavy window; otherwise all five children will die. On the windowsill there is a person waiting safely to be evacuated, who will inevitably be pushed out of the window if the glass is broken. Falling from the windowsill will certainly be fatal for that person. If Roma breaks the window, the person will be knocked off the sill and die, but the five children will be evacuated. If Roma does not break the window, the person will be saved but the children will die. Roma decides to break the window.*
   *Breaking the window is:*

7. Толя везет пять тяжело больных людей в больницу. Они в критическом состоянии и умрут, если Толя остановится по дороге. Погружая людей в машину, Толя прижал дверью толстый шнур больше метра длиной, который теперь болтается сбоку машины. Он выбрал самый быстрый путь к больнице, это узкая горная грунтовка. По пути Толя видит скалолаза, висящего на камнях возле дороги. Скалолаз в безопасности и владеет ситуацией, но если Толя проедет мимо, то болтающийся сбоку машины шнур собьет скалолаза, он упадет и разобьется. Если Толя затормозит и подождет, то скалолаз доберется до надежной опоры, где шнур не заденет его, но спасти пятерых человек уже не удастся. Если Толя не остановится, скалолаз разобьется, а пять человек будут спасены. Толя решает продолжить движение.
Продолжить движение – это:

*Tolya is driving five seriously ill people to the hospital. They are in critical condition and will die if Tolya stops on the way. While loading the patients into the car, Tolya accidentally clipped a long thick cord (over a meter) in the door so that it now dangles outside the vehicle. He chose the fastest route, a narrow mountain dirt road. Along the way Tolya sees a climber hanging on rocks beside the road. The climber is currently safe and in control, but if Tolya passes by, the dangling cord will catch the climber, who will fall and be killed. If Tolya stops and waits, the climber will reach secure footing and the cord will not endanger him, but then Tolya will not be able to get the five patients to the hospital in time and they will die. If Tolya does not stop, the climber will die and the five patients will be saved. Tolya decides to continue driving.*
*Continuing to drive is:*

8. Максим работает в больнице. Он управляет устройством, которое увеличивает или уменьшает объем определенного химического вещества в крови пациента. Если вещества слишком много или слишком мало, то пациент умирает. Максим замечает, что устройство накачало в тело пациента почти токсическую дозу вещества. Он должен немедленно остановить работу устройства, чтобы спасти жизнь пациента. Однако он видит, что пять других пациентов с токсическими дозами вещества подключены к этому же устройству, причем оно удаляет это вещество из их органов. Если Максим остановит работу устройства для спасения одного пациента, то умрут другие пять пациентов. Если Максим не остановит устройство, умрет один пациент, но пятеро выживут. Максим решает не останавливать работу устройства.
   Не остановить устройство – это:

   *Maksim works at a hospital operating a device that increases or decreases the level of a certain chemical in a patient's blood. If the level is too high or too low, the patient will die. Maksim notices that the device has pumped almost a toxic dose into one patient. He must immediately stop the device to save that patient. However, he also sees that five other patients with toxic levels are connected to the same device, and it is removing the substance from their bodies by transferring it to the first patient. If Maksim stops the device to save the one patient, the other five will die. If he does not*

*stop the device, the one patient will die and the five will survive. Maksim decides not to stop the device.*
*Not stopping the device is:*

9. Витя получил сообщение, что капитан грузового судна заразился опасным инфекционным заболеванием. Сам капитан – лишь переносчик заболевания, и имеет иммунитет, однако любой, кто вступит с ним в контакт, умрет. На судне нет пассажиров. Оно направляется на отдаленный остров, где капитан должен лично вручить груз пяти жителям острова. Капитан не знает, что является носителем заболевания, и на борту судна нет радио, чтобы предупредить капитана. Витя вылетает на вертолете, чтобы остановить судно, но уже на расстоянии видит, что корабль пристает к берегу. Однако на корабле пожар, и капитан погибнет еще до швартовки. Витя может запустить ракету в воду рядом с кораблем, и всплеск от ракеты потушит пожар и спасет капитана. Если Витя запустит ракету, капитан выживет, но заболевание перейдет на остров и убьет пять жителей. Если Витя не запустит ракету, капитан умрет, а пять островитян выживут. Витя решает выстрелить.
   Пустить ракету – это:

   *Vitya receives information that the captain of a cargo ship has contracted a dangerous infectious disease. The captain himself is merely a carrier and is immune, but anyone who comes into contact with him will die. There are no passengers on the ship. The ship is heading to a remote island where the captain is supposed to deliver cargo to five island residents. The captain does not know he is a carrier and there is no radio on board to warn him. Vitya flies by helicopter to intercept the ship and stop it, but sees from a distance that the ship is already docking. However, a fire has broken out onboard and the captain will die before the docking is complete. Vitya can fire a rocket into the water near the ship, and the resulting splash will extinguish the fire and save the captain. If Vitya fires the rocket, the captain will survive, but the disease will reach the island and kill five residents. If Vitya does not fire the rocket, the captain will die and the five islanders will survive. Vitya decides to fire the rocket.*
   *Firing the rocket is:*

10. На новой выставке аквариумов Миша видит, как посетитель поскользнулся на мокром полу, упал и сломал себе шею. Он еще жив, и, если останется неподвижным, то его могут безопасно эвакуировать медики. Однако он упал на шланг, который подает кислород пяти другим посетителям, находящимся на подводной экскурсии. Без кислорода эти пять посетителей скоро умрут. Если Миша ничего не сделает, то первого посетителя безопасно эвакуируют, но пять человек, находящихся в батискафе, умрут. Если Миша вытянет из-под него шланг, то этот посетитель умрет, но пять посетителей в батискафе получат кислород и выживут. Миша решает вытянуть кислородный шланг.
    Вытянуть шланг – это:

*At a new aquarium exhibition Misha sees a visitor slip on the wet floor, fall, and break his neck. The visitor is still alive and could be safely evacuated if left still. However, he has fallen on a hose that supplies oxygen to five other visitors on an underwater tour. Without oxygen those five visitors will soon die. If Misha does nothing, the first visitor will be safely evacuated but the five in the bathyscaphe will die. If Misha pulls the hose from under him, that visitor will die, but the five visitors in the bathyscaphe will receive oxygen and survive. Misha decides to pull the oxygen hose.*
*Pulling the hose is:*

11. Стоя рядом с рельсами, Ваня увидел, что пустой неуправляемый вагон вот-вот собьет пять человек. Рядом с Ваней находится рычаг, потянув который можно опустить перила на пешеходном мосту, чтобы один человек упал оттуда на рельсы, где его собьет вагон. Из-за этого вагон замедлится, и пять человек успеют спастись. Если Ваня потянет за рычаг, то один человек упадет на рельсы и попадет под вагон, а пять человек выживут. Если Ваня не потянет за рычаг, вагон поедет дальше, собьет пять человек, а один человек на мосту останется в живых. Ваня решает потянуть за рычаг.
Потянуть за рычаг – это:

    *Standing next to the tracks, Vanya sees an empty runaway car that is about to strike five people. Near Vanya is a lever which, if pulled, will lower the rail of a pedestrian bridge so that one person falls onto the tracks and is hit by the car. As a result, the car will slow and the five people will have time to be saved. If Vanya pulls the lever, one person will fall onto the tracks and be struck while the five will survive. If Vanya does not pull the lever, the car will continue and hit the five people, and the person on the bridge will remain alive. Vanya decides to pull the lever.*
    *Pulling the lever is:*

12. Петя – пожарник, он пытается спасти пятерых детей из горящего дома. В доме только одно окно, откуда можно безопасно эвакуировать детей, и его заклинило. Петя должен немедленно чем-то выломать это большое тяжелое окно, иначе все пять детей погибнут. Единственный достаточно большой объект – это человек, направляющийся в сторону безопасного выхода из дома. Несомненно, что если он вылетит через окно, это убьет его. Если Петя выломает окно, вытолкнув через него человека, тот умрет, но пятерых детей удастся эвакуировать. Если Петя не вытолкнет человека, то спасется этот человек, однако погибнут пять детей. Петя решает вытолкнуть человека.
Вытолкнуть человека – это:

    *Petya is a firefighter trying to save five children from a burning house. There is only one window through which the children can be safely evacuated, and it is jammed.*

*Petya must immediately force the large heavy window open, otherwise all five children will die. The only sufficiently large object available is a person who is moving toward the house's safe exit. If that person is forced out through the window, it will certainly be fatal. If Petya forces the window and in doing so pushes the person out, that person will die but the five children will be evacuated. If Petya does not push the person out, that person will survive but the five children will die. Petya decides to push the person out.*
*Pushing the person out is:*

13. Паша везет пять тяжело больных людей в больницу. Они в критическом состоянии и умрут, если Паша остановится по дороге. Погружая людей в машину, Паша прижал дверью толстый шнур больше метра длиной, который теперь болтается сбоку машины. Он выбрал самый быстрый путь к больнице, это узкая горная грунтовка. По пути Паша видит скалолаза, висящего на камнях возле дороги. Скалолаз вот-вот сорвется и разобьется насмерть, но если ему протянуть шнур, то он может выбраться и спастись.
Если Паша остановится, то скалолаз сможет использовать шнур, болтающийся сбоку машины, чтобы выбраться, однако спасти пятерых человек уже не удастся. Если Паша не остановится, один человек разобьется насмерть, а пятеро будут спасены. Паша решает продолжить движение.
Продолжить движение – это:

*Pasha is driving five critically ill people to the hospital. They will die if Pasha stops on the way. While loading them into the car, he trapped a long thick cord in the door so that it now dangles outside the vehicle. He chose the fastest, narrow mountain road. Along the way Pasha sees a climber hanging on the rocks near the road who is about to fall and die; if someone throws him the dangling cord, he could use it to get to safety. If Pasha stops, the climber could use the cord to save himself, but the five patients will not reach the hospital in time and will die. If Pasha does not stop, one person will die and the five will be saved. Pasha decides to continue driving.*
*Continuing to drive is:*

14. Стоя рядом с рельсами, Серёжа увидел, что пустой неуправляемый вагон вот-вот собьет пять человек. На пешеходном мосту над рельсами находится еще один человек, он поскользнулся и вот-вот упадет на рельсы, где его задавит вагон. Из-за этого вагон замедлит ход и пять человек успеют спастись. Рядом с Серёжей находится рычаг, потянув за который можно поднять перила моста и спасти этого человека от падения. Если Серёжа не потянет за рычаг, этот человек попадет под вагон, вагон замедлит ход и пятеро человек выживут. Если Серёжа потянет за рычаг, то вагон продолжит движение, собьет пять человек, а один человек на мосту останется в живых. Серёжа решает не тянуть за рычаг.
Не потянуть за рычаг – это:

*Standing next to the tracks, Seryozha sees an empty runaway car that is about to hit five people. On the pedestrian bridge above the tracks another person has slipped and is about to fall onto the tracks, where the car will strike him. Because of this, the car will slow and the five people will be saved. Near Seryozha there is a lever that he can pull to raise the bridge railing and thus save that person from falling. If Seryozha does not pull the lever, the person will fall under the car, the car will slow, and five people will survive. If Seryozha pulls the lever, the car will continue without slowing and will hit five people, while the person on the bridge will remain alive. Seryozha decides not to pull the lever.*
*Not pulling the lever is:*

15. Стас – пожарник, он пытается спасти пятерых детей из горящего дома. В доме только одно окно, откуда можно безопасно эвакуировать детей, и его заклинило. Стас должен немедленно выломать это большое тяжелое окно, иначе все пять детей погибнут. Под окном на земле находится человек, в безопасности ожидающий эвакуации. Если Стас выбьет стекло, оно неизбежно упадет на этого человека. Падение стекла несомненно окажется для него смертельным. Если Стас выломает окно, стекло упадет на человека, и тот погибнет, но пятерых детей удастся эвакуировать. Если Стас не выломает окно, то человека спасут, но погибнут дети. Стас решает выломать окно.
    Выломать окно – это:

    *Stas is a firefighter trying to save five children from a burning house. There is only one window through which the children can be safely evacuated, and it is jammed. Stas must immediately break this large heavy window, otherwise all five children will die. Under the window on the ground there is a person waiting to be evacuated who is in a safe position. If Stas breaks the glass, it will inevitably fall on that person. The falling glass will certainly be fatal. If Stas breaks the window, the glass will fall on the person and kill them, but the five children will be evacuated. If Stas does not break the window, the person will be saved but the children will die. Stas decides to break the window.*
    *Breaking the window is:*

16. Боря работает в больнице. Он управляет устройством, которое увеличивает или уменьшает объем определенного химического вещества в крови пациента. Если вещества слишком много или слишком мало, то пациент умирает. Боря замечает, что устройство накачало в тело пациента почти токсическую дозу вещества. Он должен немедленно отключить пациента от устройства, чтобы спасти его жизнь. Однако он видит, что пять других пациентов с токсическими дозами вещества подключены к этому же устройству. Оно удаляет вещество из их органов за счет закачивания этого вещества в организм первого пациента, и может работать, пока поддерживается этот баланс. Если Боря отключит устройство для спасения одного пациента, это устройство не сможет поддерживать баланс, и другие пять пациентов умрут. Если Боря не отключит пациента от устройства, то этот один пациент умрет, но выживут другие пять. Боря решает не отключать пациента.
Не отключить пациента от устройства – это:

*Borya works in a hospital operating a device that increases or decreases the amount of a certain chemical in a patient's blood. If the level is too much or too little, the patient dies. Borya notices the device has pumped nearly a toxic dose into one patient and must immediately disconnect that patient to save their life. However, he sees five other patients with toxic levels connected to the same device; the device is removing the substance from their bodies by transferring it into the first patient and can only continue while this balance is maintained. If Borya disconnects the patient to save them, the device will stop balancing and the other five patients will die. If Borya does not disconnect the patient, that one patient will die but the other five will survive. Borya decides not to disconnect the patient.*
*Not disconnecting the patient is:*

17. Илья управляет стрелкой на железнодорожной станции. Он видит, что пустой неуправляемый вагон едет по железной дороге с такой скоростью, что собьет любого на смерть. Вагон вот-вот собьет рабочего, нога которого застряла в стрелке. Дальше на пути находятся еще пять рабочих.
Если Илья ничего не сделает, вагон собьет одного рабочего, остановится и не заденет пятерых рабочих. Илья может перевести стрелку и освободить ногу рабочего, позволив ему отпрыгнуть с путей. Однако, освободив ногу одного рабочего, Илья даст вагону проехать и сбить пять рабочих. Илья решает не тянуть за рычаг.
Не потянуть за рычаг – это:

*Ilya operates a railway switch. He sees an empty runaway wagon traveling at a speed that would kill anyone it hits. The wagon is about to strike a worker whose foot is stuck in the switch. Further along the track there are five other workers. If Ilya does nothing, the wagon will hit one worker, stop, and not reach the five workers. Ilya can switch the track and free the trapped worker's foot, allowing him to jump clear of the tracks. However, by freeing that worker, Ilya would direct the wagon onto the other track where it would strike five workers. Ilya decides not to pull the lever.*
*Not pulling the lever is:*

18. Юра получил сообщение, что капитан грузового судна заразился опасным инфекционным заболеванием. Сам капитан – лишь переносчик заболевания, и имеет иммунитет, однако любой, кто вступит с ним в контакт, умрет. На судне нет пассажиров. Оно направляется на отдаленный остров, где капитан должен лично вручить груз пяти жителям острова. Капитан не знает, что является носителем заболевания, и на борту судна нет радио, чтобы предупредить капитана. Единственный способ предотвратить заражение жителей от капитана – взорвать судно, запустив в него ракету. Если Юра запустит ракету, капитан погибнет, а пять островитян выживут. Если Юра не запустит ракету, то капитан выживет, но умрут пять жителей острова. Юра решает пустить ракету.
Запустить ракету – это:

*Yura receives information that the captain of a cargo ship has contracted a dangerous infectious disease. The captain is only a carrier and is immune, but anyone who contacts him will die. There are no passengers on board. The ship is headed to a remote island where the captain is to personally deliver goods to five islanders. The captain does not know he is a carrier and there is no radio onboard to warn him. The only way to prevent the captain from infecting the islanders is to destroy the ship by firing a rocket at it. If Yura fires the rocket, the captain will die and the five islanders will survive. If Yura does not fire the rocket, the captain will live and the five islanders will die. Yura decides to fire the rocket.*
*Firing the rocket is:*